%% file: main.tex
\documentclass{resources/aimc2026}
\input{resources/preamble}
\begin{document}

\twocolumn[%
\maketitle
\begin{abstract}
Quadratic difference tones (QDTs) are a species of auditory distortion product in which a ``phantom'' pure tone, absent from the acoustic signal, is clearly audible to listeners. Exploiting this phenomenon, one can synthesize harmonically rich tones for musical purposes, a technique called Quadratic Difference Tone Spectrum (QDTS) synthesis. Previous works have introduced numerical methods to synthesize QDTS based on the distortion function, which links a target QDTS and an overtone-structured carrier signal. While accurate, these methods were stochastic and discontinuous, making them difficult to control for musical purposes and effectively limiting them to stationary signals. This paper proposes a neural network-based approach that learns an approximate inverse of the distortion mapping in an autoencoder-like configuration, producing a continuous approximation that addresses prior limitations. Experimental results show that, although slightly less numerically precise, the method is sufficient for perceptual and musical applications. We also implement a real-time version in Max and evaluate its performance. Various sound examples demonstrate its expressive and musical potential. The source code, audio examples, tutorials, and software accompanying this work are available at \href{https://cordutie.github.io/projects/qdts.html}{https://cordutie.github.io/projects/qdts.html}.
\end{abstract}
]

\aimcnotice

\section{Introduction}
\label{sec:introduction}
Quadratic difference tones (QDTs) are one of a family of so-called auditory distortion products (ADPs), sound sources that appear in the presence of particular acoustic stimuli but that are absent from the physical sound signal (i.e., they do not appear in spectrogram analyses). Historically known as difference and combination tones, ADPs are today best understood as subjective sounds generated by nonlinearities in the active components of the cochlea, specifically the outer hair cells and their interaction with the basilar membrane \citep{Johnsen1983}. 
They are typical of healthy hearing and their testing has become a common diagnostic tool for identifying hearing disorders \citep{kemp1978stimulated, Johnsen1983, Abdala2001}.

In hearing studies, ADPs are typically evoked by presenting two primary sinusoidal signals, $f_1$ and $f_2$, and asking subjects to report the perceived sonority. In these conditions, two ADPs are particularly salient: the QDT that is the subject of this study, and the lower cubic difference tone (LCDT). The LCDT occurs at $2f_1 - f_2$, obeys cubic nonlinearity, and its amplitude is highly sensitive to the frequency ratio of the primaries. It also exhibits the highest level when recorded as a DPOAE in the ear canal. The QDT occurs at $f_2 - f_1$, is governed by square-law distortion, and has a lower DPOAE level. It is less sensitive to ratio than the LCDT, and thus retains its amplitude across wider variations in interval size. Additionally, the QDT is easier to perceptually segregate than the LCDT \citep{dewey, zwicker, plomp}.

QDTs have a rich musical history dating back to their discovery by violinist Giuseppe Tartini in 1754 \citep{nelson2003violin}. Their use ranges from the transient ``ghost'' tones elicited by improvisers like John Butcher and Evan Parker to the experimental drone works of Maryanne Amacher, Phill Niblock, and Catherine Christer Hennix. A primary challenge of deploying QDTs for music synthesis has been the sound pressure levels required for them to be audible. Using a two-sinusoid stimulus requires the gain to be at levels that are uncomfortable for most listeners.

To ameliorate this problem, \citeauthor{Hawor11} (\citeyear{Hawor11}) presented a method utilizing a complex of pure tones, building upon previously established psychoacoustic findings \citep{pressnitzer2001distortion}. By arranging the frequencies so that each consecutive pair of acoustic sinusoids produces the identical QDT, this configuration boosts the overall distortion gain while simultaneously generating harmonic components of the primary QDT. Furthermore, increasing the number of acoustic tones and spreading them over a wider frequency range allows the subjective level of the acoustic stimulus to be reduced, which greatly diminishes listener fatigue. This technique, termed Quadratic Difference Tone Spectrum (QDTS) synthesis, is the focus of this article. Formalizing the approach, \citeauthor{kendall} (\citeyear{kendall}) proposed models for QDTS synthesis,
which afford unprecedented control over the timbre of the resulting auditory illusion.

This shift from raw phenomenon to controllable synthesis model has spurred a recent wave of integration into contemporary electronic music and sound art. The precise architectural manipulation of QDTS has been featured in recent electroacoustic releases and installations, exemplifying the impact of the technique on current experimental practices. Notable applications include the microtonal inner ear orchestrations of Marcin Pietruszewski \citep{Pietruszewski2024}, procedural and iterative resynthesis deployments by \citeauthor{Hecker2025} (\citeyear{Hecker2025}), and the dissociative synthetic vocal works of \citeauthor{Francis2025} (\citeyear{Francis2025}).

Despite these creative successes, existing algorithmic implementations of QDTS synthesis suffer from severe limitations that have constrained their musical use. The original symbolic method of \citeauthor{kendall} (\citeyear{kendall}) was bottlenecked at four harmonics, beyond which closed-form solutions are not generally available. The numerical extension introduced by \citeauthor{ICMC23} (\citeyear{ICMC23, cmj24}) lifted this restriction by employing a stochastic Newton-Raphson solver capable of handling an arbitrary number of harmonics, but in doing so introduced a critical continuity problem. Because the underlying inverse problem is generally multi-valued, the solver selects a different local inverse at each time step with no mechanism to enforce consistency between successive solutions. When the target spectrum varies in time, as it must in any musically interesting context, the recovered carrier complex jumps between distant solution branches, producing audible discontinuities in the synthesized signal. In practice, this restricted prior tools to stationary or near-stationary targets, severely limiting their expressive range.

The present paper proposes a neural network-based solution to the continuity problem that renders dynamic QDTS synthesis viable as a robust, real-time tool for contemporary composers. The central idea is to replace stochastic root-finding with a learned, deterministic mapping from target spectra to carrier complexes. Concretely, we frame the inverse of the distortion function as the encoder of an autoencoder-like configuration in which the (analytically known) distortion function itself acts as a fixed decoder. Continuity is then a structural property of the learned map rather than a quantity that must be enforced post-hoc. To respect the degree-2 homogeneity of the distortion function, we factor the network architecture so that scaling behavior is exact by construction, leaving the network only to learn the angular component of the inverse on the unit sphere. Together, these design choices yield a solver that is continuous, deterministic, and inexpensive to evaluate.

The contributions of this paper are threefold. First, we introduce a continuous, neural approximation to the QDTS inverse problem, trained through a curriculum-style procedure that mitigates the multiplicity of local inverses by progressively expanding the sampling domain. Second, we provide a real-time Max external and a companion patch, making the method directly accessible to composers in a familiar environment. Third, we present a quantitative evaluation of the proposed solver against the prior Newton-Raphson baseline along three axes: reconstruction accuracy, control smoothness, and computational efficiency. Additionally, we complement these results with a set of sound examples demonstrating previously unattainable forms of expressive control, including continuous timbral morphing, resynthesis of time-varying monophonic audio, and explicit manipulation of carrier phase as a compositional parameter.

The remainder of the paper is structured as follows. Section~\ref{sec:background} develops the mathematical formulation of the QDTS model and analyzes the perceptual constraints, feasibility issues, and continuity problem inherited from prior work. Section~\ref{sec:model} introduces our neural solver, including its homogeneity-preserving architecture and curriculum-style training procedure. Section~\ref{sec:max} describes the real-time Max implementation and presents the quantitative benchmarks. Section~\ref{sec:others} discusses the accompanying sound examples, with attention to resynthesis and phase-dependent regimes. Section~\ref{sec:conclusions} concludes the paper and Section~\ref{sec:future_work} outlines directions for future work, with particular emphasis on the open problem of extending the framework to cubic difference tones.

\section{Background}\label{sec:background}

In this section we introduce the QDTS model using a formulation consistent with our proposed approach. We also discuss the intrinsic limitations and problems with the model, including perceptual constraints, mathematical feasibility, and controllability issues.

\subsection{Quadratic Difference Tone Spectra in a Nutshell}
The distribution of amplitudes of a Quadratic Difference Tone Spectrum (QDTS) has been theorized by \cite{kendall} and further explored by \citeauthor{ICMC23} (\citeyear{ICMC23, cmj24}). For convenience in our formulation, we give a brief summary of the functional formulation proposed in the latter. 

\begin{figure}
\begin{center}
\includegraphics[width=\columnwidth]{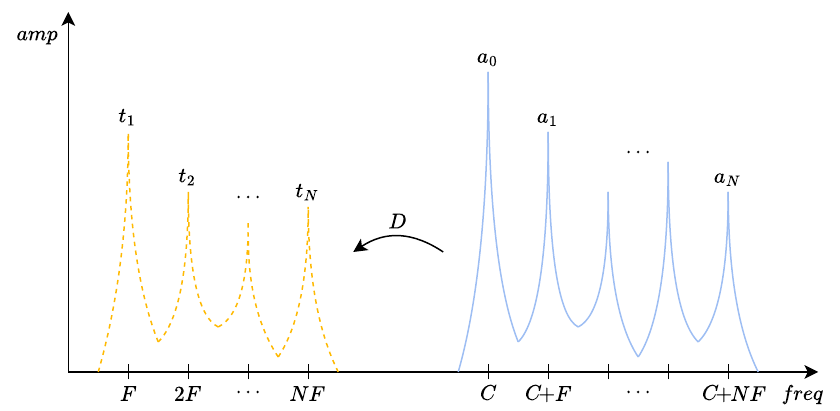}
\caption{Illustrative frequency-amplitude representation of the QDTS model. On the right, the distribution of amplitudes of the carrier complex of pure tones is shown in cyan. On the left, the distribution of amplitudes of the target complex, evoked as a distortion product by the carrier, is shown dashed and in orange. The distortion function $D$ is added to show how these signals are connected.}
\label{fig:graphical_representation}
\end{center}
\end{figure}

In the QDTS model, there are two signals: one acoustic signal representing the \textit{carrier} complex of sinusoids, and one representing the perceptual auditory distortion product evoked by the carrier, acting as the \textit{target} complex of harmonically distributed sinusoids. In this framework, the \textit{carrier} complex is comprised of sinusoids of frequencies $C, C+F, C+2F, \dots, C+NF$, where $C,F>0$ and $N$ is an integer; and the \textit{target} QDTS generated by such complex is then comprised of $N$ pure tones of frequencies $F, 2F, \dots, NF$. Moreover, if the amplitudes of the carrier complex of sinusoids are $a_0, \dots, a_N$, the amplitudes of the harmonics of the QDTS, respectively $t_1, \dots, t_N$, are the result of the application of the \textit{distortion function} $D:\mathbb{R}^{N+1}\rightarrow\mathbb{R}^N$ on the carrier distribution of amplitudes:
\begin{equation}
D(a_0,\dots,a_N)=(t_1,\dots,t_N),
\end{equation} 
where each target amplitude is computed by
\begin{align}
\label{eq:system}
t_1    & = a_0 a_N \nonumber\\
t_2    & = a_0 a_{N-1} + a_1 a_N \nonumber\\
\vdots & \hspace{15mm}\vdots \\
t_{N-1}    & = a_0 a_2 + a_1 a_3 + \dots +a_{N-2} a_{N} \nonumber\\
t_N    & = a_0 a_1 + a_1 a_2 + \dots +a_{N-2} a_{N-1}+ a_{N-1} a_N \nonumber 
\end{align}
See a graphic representation of this setup in Figure~\ref{fig:graphical_representation}.

Hence in theory, by solving the equation $D(a)=t$ for a fixed vector $t$, one can determine the carrier needed to generate any given target harmonic signal. This is what gives this model its sound synthesis appeal.

This model is not flawless and is subject to the same perceptual constraints as the underlying QDTs. These sounds are known to vary in perceived strength depending on the frequency range and the ratios of the carrier signal. The audibility of the QDT depends strongly on the intensity of the acoustic tones: a level above 50 dB SPL is required for the component to be heard, and its amplitude then increases proportionally with the primaries, such that for every 1 dB increase, the QDT grows by 2 dB \citep{hartmann2004signals}.

\subsection{Feasibility and Workarounds}\label{subsec:feasibility}
Beyond the perceptual constraints of QDTS, finding a definitive mathematical solution to Equation \eqref{eq:system} presents significant challenges. Solving $D(a)=t$ requires finding the roots of a system of multivariate polynomials. While symbolic computation yields exact solutions for systems where $N \leq 4$ \citep{kendall}, only the existence of complex solutions has been proven for orders $N=5$ and $N=6$ (\citeauthor{ICMC23} \citeyear{ICMC23}; \citeyear{cmj24}). Furthermore, simple examples lacking any real solution can easily be found for $N \geq 5$, making the problem analytically unsolvable in the general case.

Previous research (\citeauthor{ICMC23} \citeyear{ICMC23}; \citeyear{cmj24}) bypassed this limitation by employing a stochastic numerical approach. By applying the Newton--Raphson algorithm, an approximate solution could be calculated. If the algorithm failed to converge within a set number of iterations, a microscopic random perturbation was added to the target distribution, and the process was restarted. Even though there was no theoretical guarantee for an algorithm of this nature to converge, empirical testing demonstrated that this approach yielded an approximate solution that is still perceptually relevant in over 99\% of cases. Ultimately, this approach proved that despite the lack of formal mathematical certainty, computing approximate solutions remains a highly feasible workaround for the problem.

\subsection{The Continuity Problem}\label{subsec:continuity}
The distortion function $D$, like any multi-variable $C^{\infty}$ function, has the potential to admit many local inverses in different parts of its co-domain. In practice, this implies that given a target spectrum, the system of equations \eqref{eq:system} generally may admit multiple real solutions, and therefore an infinite set of approximate solutions.

This multiplicity becomes problematic when the target spectrum evolves over time. Numerical and stochastic solvers, such as the method introduced by \citeauthor{ICMC23} (\citeyear{ICMC23,cmj24}), effectively select one of these local inverses at each step, but provide no mechanism to enforce consistency between successive solutions. As a result, the recovered carrier complex may jump between distant solution branches, producing abrupt changes in the carrier signal. These discontinuities manifest as audible glitches or artifacts, and up to this contribution, remained as an open problem.

\section{Learned Inverse Distortion Function}
\label{sec:model}
In essence, solving the QDTS model amounts to finding a map that transforms target spectra into carrier complexes. We call this map the \textit{synthesis function} $S:\mathbb{R}^{N}\rightarrow\mathbb{R}^{N+1}$. Ideally, $S$ should act as a left inverse of the distortion function, meaning that it produces outputs whose distortion matches the target spectrum:
\begin{equation}\label{eq:inverse}
D(S(t)) = t.
\end{equation}
Moreover, in order to address continuity issues, $S$ must also be continuous.

\begin{figure}
\begin{center}
\includegraphics[width=\columnwidth]{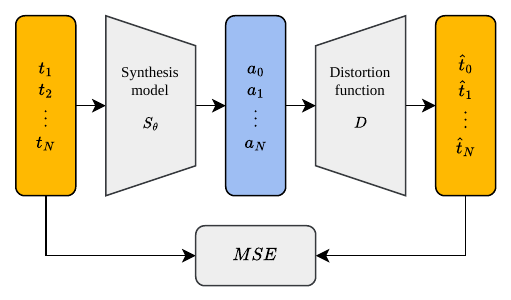}
\caption{Autoencoder-like architecture. The distortion function works as a decoder, while the synthesis function is learned.}
\label{fig:autoencoder}
\end{center}
\end{figure}

As discussed in Subsection~\ref{subsec:feasibility}, the relation \eqref{eq:inverse} may not always be achievable; however, the relaxation $D(S(t)) \approx t$ is feasible and still perceptually relevant.

In this section, we propose to learn a map $S = S_\theta$ with these properties using a shallow neural network, and to enforce such properties through simple architectural and training choices.

\subsection{Architecture}
The distortion function is homogeneous of degree $2$, i.e., for $\lambda\in\mathbb{R}$ and $a\in\mathbb{R}^{N+1}$,
\begin{equation}
D(\lambda a) = \lambda^2 D(a).
\end{equation}
It follows that if $S$ is a left inverse of $D$, it must satisfy the corresponding inverse relation
\begin{equation}
\label{eq:inverse_homogeneous}
S(\lambda t)=\sqrt{\lambda}\,S(t),
\end{equation}
for $t\in\mathbb{R}^{N}$ and $\lambda>0$.

We enforce this property by construction:
\begin{equation}\label{eq:construction}
S_\theta(t) = \sqrt{\|t\|}\, s_\theta\left(\frac{t}{\|t\|}\right),
\end{equation}
where $t\neq 0$ and $s_\theta$ is a neural network.

This construction not only ensures Equation~\eqref{eq:inverse_homogeneous}, but it also guarantees continuity, improves training stability by operating exclusively on normalized vectors, and allows the function to extend naturally to the full domain once trained on the $N$-dimensional sphere.

Given the low complexity of the problem and the lack of an obvious relationship between input dimensions, a natural baseline is a shallow Multi-Layer Perceptron (MLP) directly modeling $s_\theta$. Adopting this baseline, empirical testing revealed that a three-layer MLP yielded the best results.

\subsection{Training}
This formulation can be interpreted as an autoencoder-like architecture in which the distortion function $D$ acts as a fixed decoder mapping acoustic signals to psychoacoustic phenomena, while $S_\theta$ serves as a learned encoder performing the inverse mapping. The model is trained by minimizing the reconstruction loss
\begin{equation}\label{eq:loss}
    \mathcal{L}(t) = \| D(S_\theta(t)) - t \|^2.
\end{equation}
See an illustration of this autoencoder-like approach in Figure~\ref{fig:autoencoder}.

As noted earlier, some regions of the codomain of $D$ may admit multiple inverses, which can complicate training. To mitigate this, we train on random batches sampled from spheres of increasing radius. This encourages the model to first learn a local inverse and then extend it smoothly to larger domains.

Specifically, we begin with $10{,}000$ batches of $512$ random target vectors sampled from a sphere centered at $(0.5, \dots, 0.5)$ with radius $0.1$, and progressively increase the radius in steps of $0.1$ up to $1$.

Due to non-convexity and the presence of multiple possible inverses, we perform $8$ runs with different initializations for each $N\in\{5,\dots,16\}$ and another $8$ runs using a regularizer that makes $a_0\approx t_1$, a constraint also used by \citeauthor{ICMC23} (\citeyear{ICMC23, cmj24}) that helps localize the solutions. Although this procedure provides no formal guarantees, it consistently yields accurate and continuous approximate inverses as shown in Section~\ref{sec:max:gen_quality}.

\section{Implementation in Max and Benchmarks}
\label{sec:max}
In this section, we introduce a real-time implementation of our neural network-based QDTS synthesizer, alongside a companion patch designed to seamlessly integrate this technique into compositional workflows. To better understand the performance and quality of the developed external, we also conduct a series of quantitative experiments. The objective of these experiments is to evaluate the quality of the reconstructed target spectra, the smoothness of the output under continuous parameter changes, and the computational efficiency of our implementation in a real-time setting. 

\subsection{Max Implementation}
To allow composers to interact with the trained models in a familiar environment, we developed a custom Max external called \texttt{qdts.solver\_nn}, as well as a companion Max patch \citep{max9, max1990}.\footnote{The external, patch, and the source codes are openly accessible from \href{https://cordutie.github.io/projects/qdts.html}{https://cordutie.github.io/projects/qdts.html}.}

The external uses \cite{onnxruntime} to run inference on the pre-trained models, exported in the ONNX format \citep{bai2019}. This approach results in a self-contained package that is straightforward to install in a manner with which Max users are familiar.

\begin{figure}[h]
\begin{center}
\includegraphics[width=\columnwidth]{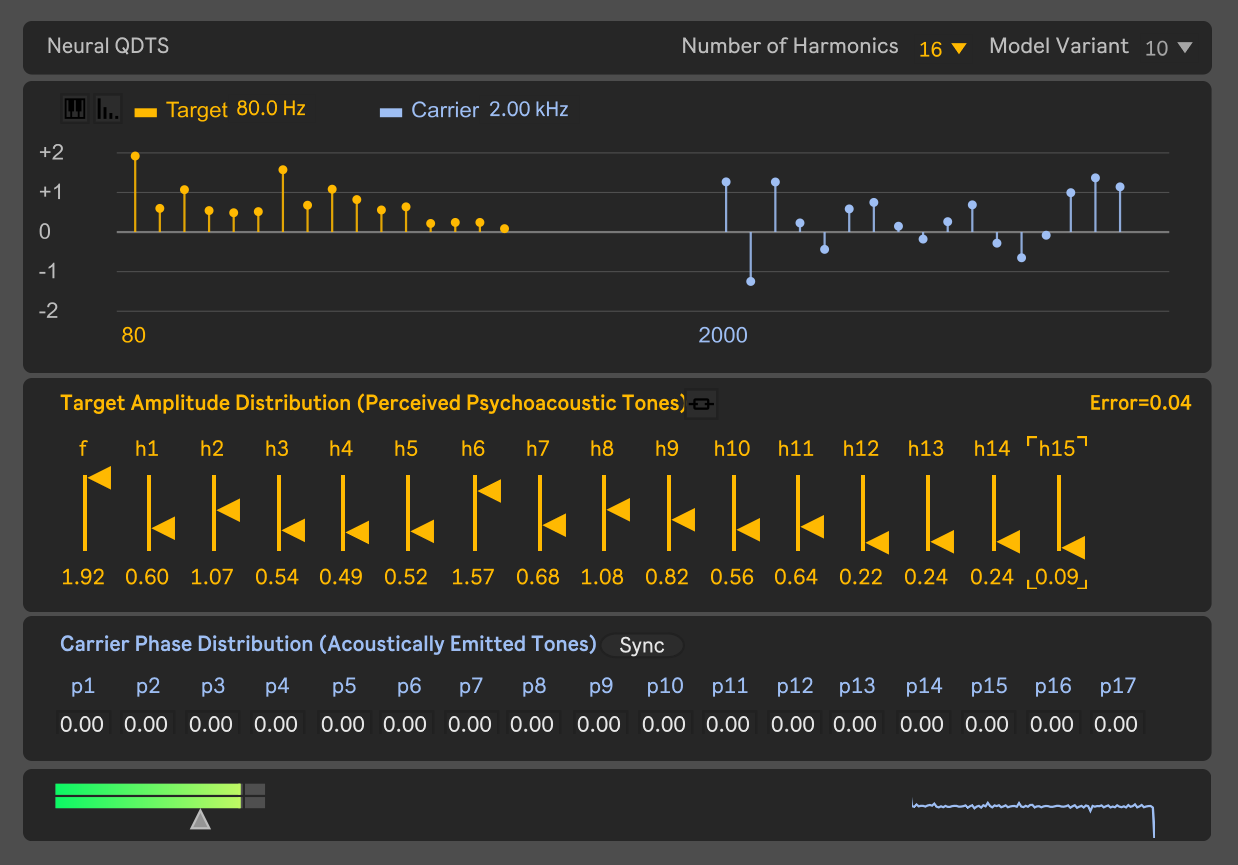}
\caption{A snapshot of the developed Max patch to allow for easy interaction with the \texttt{qdts.solver\_nn} external employing the trained models.}
\label{fig:max_patch}
\end{center}
\end{figure}

The patch (see Figure~\ref{fig:max_patch}) allows the user to specify the target and carrier frequencies and the number of harmonics, and to interactively adjust the target amplitude distribution in real time. Based on the configuration, \texttt{qdts.solver\_nn} automatically selects a compatible model and converts the target amplitudes to the corresponding carrier amplitudes, which are then passed to a custom additive synthesizer to auralize the resulting spectrum. Moreover, the user can choose any of the $16$ models trained for each case, giving further possibilities to the timbre of the acoustic signal while generating the same QDTS. 

\subsection{Generation Quality}\label{sec:max:gen_quality}
To assess the quality of the generations, we evaluated the reconstruction error of the deployed models across all supported target spectrum sizes $N \in \{5, \dots, 16\}$ using 10{,}000 random target amplitude vectors drawn uniformly from $[0, 1]^N$. For each input, reconstruction error is measured by the same loss function used during training, $\mathcal{L}$ in Equation~\eqref{eq:loss}, to quantify how closely the distortion products of the predicted carrier match the target spectrum.

\begin{figure}[h]
\begin{center}
\includegraphics[width=\columnwidth]{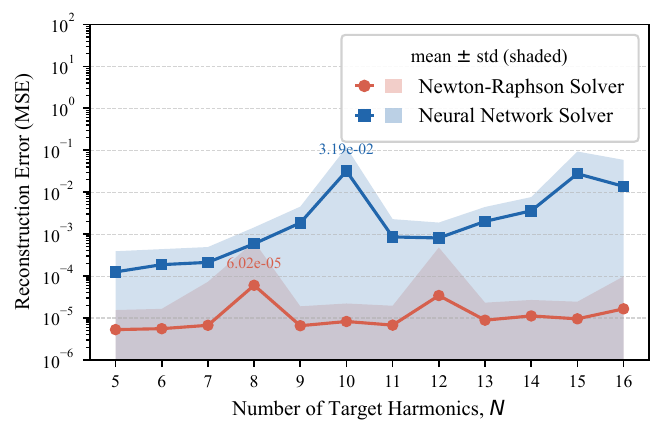}
\caption{Reconstruction error as a function of $N$ for the neural network solver and the Newton--Raphson solver. For each $N$, the results are aggregated across all 16 model variants.}
\label{fig:quality}
\end{center}
\end{figure}

For reference, we also compare against the Newton--Raphson implementation of \citeauthor{ICMC23} (\citeyear{ICMC23, cmj24}), evaluating the same metric. Figure~\ref{fig:quality} shows the mean and standard deviation of $\mathcal{L}$ as a function of $N$ for both solvers.

As shown in Figure~\ref{fig:quality}, the Newton--Raphson solver achieves lower reconstruction error across all values of $N$. Nevertheless, the neural network solver remains well within an acceptable range, with mean errors below $0.04$ across all supported spectrum sizes, confirming that it reliably approximates the inverse of the distortion function throughout the supported range. The narrow variance band indicates that performance is consistent across model variants and random inputs, suggesting that the deployed models yield stable solutions.

\subsection{Control Smoothness}\label{sec:max:smoothness}

To evaluate whether \texttt{qdts.solver\_nn} produces smooth outputs under continuous parameter changes, we measured how closely the model's response to a linearly interpolated input follows a linear interpolation of the endpoint outputs. For each target spectrum size $N \in \{5, \dots, 16\}$, we generated 10{,}000 random endpoint pairs $\mathbf{A} \sim \mathcal{U}[0,1]^N$ and $\mathbf{B} = 1 - \mathbf{A}$, and queried the model at 11 evenly spaced interpolation points $\mathbf{T}_\alpha = (1-\alpha)\mathbf{A} + \alpha\mathbf{B}$, $\alpha \in \{0, 0.1, \ldots, 1.0\}$, across all 16 model variants. At each $\alpha$, smoothness is quantified by the normalized output distance

\begin{equation}
    r(\alpha) = \frac{\lVert S_\theta(\mathbf{T}_\alpha) - S_\theta(\mathbf{A}) \rVert}
                     {\lVert S_\theta(\mathbf{B})         - S_\theta(\mathbf{A}) \rVert},
\end{equation}

which equals $\alpha$ for a perfectly linear model. 

\begin{figure}[h]
\begin{center}
\includegraphics[width=\columnwidth]{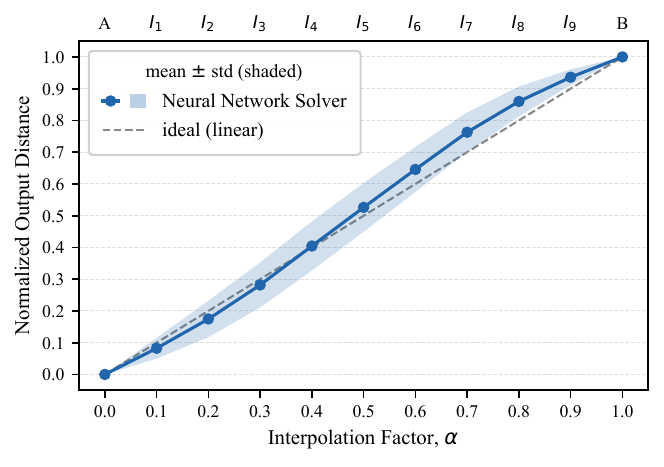}
\caption{Normalized output distance $r(\alpha)$ of the neural network solver at 11 interpolation points between two random endpoint spectra, pooled across all target spectrum sizes $N \in \{5, \dots, 16\}$ and all 16 model variants. The dashed diagonal represents the ideal linear response $r(\alpha) = \alpha$.}
\label{fig:smoothness}
\end{center}
\end{figure}

Figure~\ref{fig:smoothness} shows the distribution of $r(\alpha)$ pooled across all $N$ and model variants. This analysis is specific to the neural network solver; an equivalent evaluation of the Newton--Raphson solver is not meaningful, as it selects solutions independently at each time step with no mechanism to enforce consistency between successive outputs, and would therefore produce arbitrary $r(\alpha)$ values regardless of input smoothness.

As shown in Figure~\ref{fig:smoothness}, \texttt{qdts.solver\_nn} produces smooth, nearly linear outputs under continuous input interpolation. The normalized output distance $r(\alpha)$ tracks the ideal diagonal closely across all interpolation points, indicating that the carrier tone amplitudes transition continuously and predictably as the target spectrum is varied, with only a small degree of nonlinearity. The consistently narrow variance band across all values of $N$ and model variants confirms that this smooth behavior is stable and not specific to particular inputs or configurations.

\subsection{Computational Efficiency}\label{sec:max:comp_eff}
\begin{figure}[h]
\begin{center}
\includegraphics[width=\columnwidth]{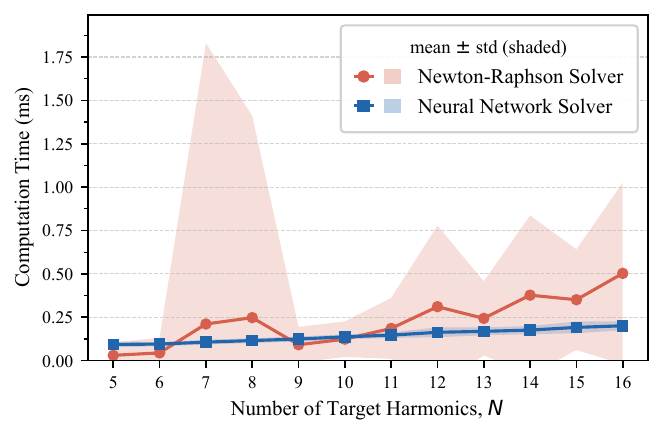}
\caption{Computation time of Newton--Raphson solver and neural network solver. Neural network solver results are aggregated across all 16 model variants.}
\label{fig:performance_benchmark}
\end{center}
\end{figure}

The computational efficiency of \texttt{qdts.solver\_nn} was evaluated by measuring the execution time of the Max external\footnote{Measured on an Apple M4 Pro.} across all supported target spectrum sizes $N \in \{5,\dots, 16\}$ using 10{,}000 random amplitude sequences. For reference, the same benchmark was applied to the Newton--Raphson implementation of \citeauthor{ICMC23} (\citeyear{ICMC23, cmj24}). Figure~\ref{fig:performance_benchmark} summarizes the results.

As shown in Figure~\ref{fig:performance_benchmark}, \texttt{qdts.solver\_nn} achieves a smooth, predictable increase in computation time with~$N$, whereas the Newton--Raphson implementation exhibits higher and more irregular times across all sizes. The stochastic nature of the Newton--Raphson restart mechanism results in high variance, particularly for ill-conditioned inputs that require many restarts before convergence. In contrast, \texttt{qdts.solver\_nn} executes a fixed sequence of neural network operations regardless of the input, yielding near-constant variance across all values of~$N$. Notably, the computation time of \texttt{qdts.solver\_nn} remains below $0.25$~ms for all values of~$N$, demonstrating its suitability for real-time audio applications.

\section{Sound Examples}
\label{sec:others}
This section presents a series of sound examples demonstrating the neural network-based implementation of QDTS synthesis. The examples are designed to illustrate both the robustness of the learned model across different pitch conditions and its behavior in resynthesis, interaction with acoustic signals, and phase-dependent regimes. All audio examples are available in the supplementary material repository\footnote{\href{https://cordutie.github.io/projects/qdts.html}{https://cordutie.github.io/projects/qdts.html}}. In all cases, synthesis is performed using our proposed neural network solver. To achieve the spatial and perceptual effects typical of QDTs, listeners are advised to use loudspeakers at moderate to high playback levels rather than headphones.

\subsection{Basic Synthesis Examples}
In this first set of examples, we demonstrate stationary pitch synthesis using our method. The objective is to evaluate the ability of the neural solver to generate stable QDTS structures while allowing continuous control over its harmonic content, a previously impossible feature.

Each example consists of a fixed-pitch synthetic tone generated via additive synthesis, which is then used as a target representation for the neural solver. The model outputs a corresponding carrier complex, with a different number of harmonics in each case. 

We repeat this procedure for three distinct fundamental frequencies using three independently trained models, demonstrating that the learned mapping generalizes across pitch space and does not depend on a single configuration. The configurations used in each example are detailed below, where $\mathcal{M}_m^n$ denotes the $m$-th model variant trained for the $N=n$ harmonic case:

\begin{itemize}
    \item Example A: $9$ harmonics with fundamental pitch $F = 73.4$~Hz, using carrier $C=2$~kHz, and neural network model $\mathcal{M}_2^9$.
    \item Example B: $10$ harmonics with fundamental pitch $F = 98.0$~Hz, using carrier $C=2.5$~kHz, and neural network model $\mathcal{M}_7^{10}$.
    \item Example C: $14$ harmonics with varying fundamental pitch, using carrier $C=2.5$~kHz, and neural network model $\mathcal{M}_{11}^{14}$.
\end{itemize}

Across all cases, our model output preserves stable perceptual pitch while exhibiting controlled redistribution of harmonic energy through the learned synthesis function.

\subsection{Resynthesis Examples}
This subsection presents QDTS resynthesis of real-world monophonic audio recordings using joint pitch tracking and harmonic distribution estimation. Although previous models have been applied to similar tasks, the lack of continuity control forced them to rely on constant harmonic distributions, a strong constraint that is lifted by our continuous model.

We first exemplify this approach using the same recording used by \citeauthor{ICMC23} (\citeyear{ICMC23}) for the purpose of comparison and add a second sound example. Here, a pitch estimator and harmonic tracking module estimate time-varying fundamental frequencies and spectral envelopes, which are then fed into the neural solver to reconstruct a corresponding QDTS representation. While demonstrated here in an offline setting, the pipeline is equally applicable in a real-time context, subject to the latency constraints of frame-based analysis.

\subsection{Interactions Between Distortion Products and Acoustic Signals}
This set of examples explores perceptual interactions between QDTS-generated distortion products and externally introduced acoustic sinusoids. The goal is to examine beating effects and interference phenomena arising when external tones are positioned near predicted distortion frequencies.

A long-form interactive example demonstrates this by introducing an additional acoustic sinusoid near the frequency of each of the first three harmonics. The listener is encouraged to notice how, despite the absence of acoustic interactions between the physical signals, audible beating occurs due to interference with the perceptual distortion products generated by our model.

\subsection{Phase Interactions}
The final set of examples investigates the role of carrier phase configuration in QDTS synthesis. While harmonic magnitudes remain fixed, variations in phase relationships of the carrier complex significantly affect the resulting distortion perceptibility.

A stationary pitch is used throughout, while the phase vector of the carrier is systematically modified. In particular, configurations are selected that either enhance or suppress the perceptual prominence of the distortion products.

\begin{itemize}
    \item Phase configuration A: Enhanced QDTS emergence.
    \item Phase configuration B: Suppressed QDTS emergence.
    \item Phase sweep experiment (continuous morphing).
\end{itemize}

The perceptual prominence of the distortion products is highly dependent on the phase configuration of the carrier signal. Users should therefore be aware that identical magnitude targets may yield markedly different perceptual outcomes depending on the carrier phase. For this reason, the Max patch exposes direct control over individual phase parameters, allowing the performer to explore and exploit these interactions in real time.

\section{Conclusions}
\label{sec:conclusions}
This paper has presented a neural network-based approach to the synthesis of Quadratic Difference Tone Spectra that resolves the continuity limitations of prior numerical solvers and renders dynamic QDTS synthesis viable for real-time musical use. By framing the inverse of the distortion function as the encoder of an autoencoder-like configuration, with the analytically known distortion function acting as a fixed decoder, and by enforcing the degree-2 homogeneity of the underlying mapping at the architectural level, we obtained a continuous, deterministic approximate inverse that can be evaluated cheaply at audio-rate control timescales.

Our quantitative benchmarks indicate that the neural solver trades a modest amount of numerical accuracy for a set of properties that prior solvers could not provide simultaneously. Compared with the stochastic Newton--Raphson baseline of \citeauthor{ICMC23} (\citeyear{ICMC23, cmj24}), the proposed model exhibits slightly higher reconstruction error, yet mean errors remain below 0.04 across all supported harmonics, an acceptable range for perceptual listening. In exchange, the neural solver produces near-linear responses to continuous parameter interpolation, and it executes comfortably within the budget of real-time audio applications. Crucially, both properties are structural rather than incidental: continuity follows from the architecture, and computational cost is independent of the input.

These properties are reflected in the accompanying sound examples, which include stable stationary synthesis across multiple fundamentals and trained models, resynthesis of time-varying monophonic recordings with continuously evolving harmonic envelopes, controlled interactions between predicted distortion frequencies and externally introduced acoustic sinusoids, and phase-dependent regimes in which identical magnitude targets yield markedly different perceptual outcomes. Several of these scenarios, in particular continuous timbral morphing and time-varying resynthesis with unconstrained harmonic distributions, were not practically achievable with prior solvers. The Max external \texttt{qdts.solver\_nn} and its companion patch make these capabilities directly available to composers in a familiar environment.


\section{Future Work}
\label{sec:future_work}
Several directions for future work follow naturally from the framework presented here. The most demanding concerns Cubic Difference Tones (CDTs), and the question of whether a coherent ``Cubic Difference Tone Spectra'' (CDTS) framework can be formulated at all. CDTs, and in particular the lower CDT at $2f_1 - f_2$, are perceptually richer than QDTs in several respects: they are typically more salient at moderate stimulus levels, they are the dominant distortion product measured in clinical distortion product otoacoustic emission (DPOAE) protocols, and they exhibit a more pronounced phenomenological separation from the acoustic primaries. From a compositional standpoint, this makes them an attractive target; from a modeling standpoint, however, the problem differs from the quadratic case in ways that prevent a direct transposition of the methodology presented here.

Three obstacles stand out. First, the underlying nonlinearity is cubic rather than quadratic, so the analogue of the distortion function $D$ is homogeneous of degree three. While this is relatively simple to enforce architecturally, it raises the polynomial degree of the system relating carrier and target amplitudes, increasing the multiplicity of local inverses. Second, and more fundamentally, the perceptual amplitude of the LCDT depends strongly and non-monotonically on the frequency ratio $f_2/f_1$ of the primaries \citep{plomp,zwicker}. The QDTS framework exploits the fact that the quadratic component is approximately invariant to interval size and adds linearly across components of a harmonic complex with constant difference frequencies; no comparable invariance holds for the cubic component, and any CDTS distortion function would have to incorporate a frequency-ratio-dependent gain term whose form is not yet well established. Third, the LCDT is more sensitive than the QDT to absolute level, primary level imbalance, and individual cochlear variability, weakening the population-level validity of any synthesis target. 

Two lower-risk directions concern the QDTS framework itself. The current solver inverts only the magnitude relationship between carrier and target, but as noted in Section~\ref{sec:others}, the perceptual prominence of QDTs depends substantially on carrier phase. Incorporating phase as a controllable target dimension, either by training joint magnitude--phase inverses or by learning auxiliary maps that select phase configurations optimizing perceptual prominence, would tighten the link between the model's mathematical targets and the listener's experience. Separately, the training procedure used here is deliberately simple, and more sophisticated schemes, including approaches that explicitly account for the multiplicity of local inverses by guiding the network toward a designated branch, could further improve reconstruction accuracy without compromising the structural properties on which the present method relies.

\section*{Acknowledgements}

This research was supported by the ANID Fondecyt Regular Grant \#1230926, funded by ANID Millenium Nucleus NCS2025-12 ANIMUPA, Government of Chile, and the project "Cátedra en IA y Música" (TSI-100929-2023-1), funded by the Secretaría de Estado de Digitalización e Inteligencia Artificial, the European Union-Next Generation EU funds and BMAT Music Innovators.

We are profoundly grateful to Marcin Pietruszewski for his support and interest in our research. His pioneering work has continually inspired this project. Furthermore, his artistic practice and research, including his SuperCollider ports\footnote{\href{https://github.com/mpietrus00/sc-qdts}{https://github.com/mpietrus00/sc-qdts}}, have served as a pillar in the dissemination of these niche but fascinating synthesis techniques to a broader audience.

\bibliographystyle{apalike}
\bibliography{resources/references}

\end{document}

%% file: resources/preamble.tex
\usepackage[utf8]{inputenc} 
\usepackage[T1]{fontenc}    
\usepackage{hyperref}       
\hypersetup{
  colorlinks = true,  
  urlcolor   = blue,  
  linkcolor  = black, 
  citecolor  = black  
}
\usepackage{url}            
\usepackage{booktabs}       
\usepackage{amsfonts, amsmath}       
\usepackage{nicefrac}       
\usepackage{microtype}      
\usepackage{graphicx}       
\usepackage{cleveref}       

\title{Learned Continuous Synthesis of\\Quadratic Difference Tone Spectra}

\author{
  Esteban Gutiérrez \\
  Department of Information and\\
  Communications Technologies \\
  Universitat Pompeu Fabra \\
  \texttt{esteban.gutierrezc@upf.edu} \\
  \And
  Behzad Haki \\
  Independent Researcher \\
  Barcelona, Spain\\
  \texttt{behzad.haki@upf.edu} \\
  \And
  Christopher Haworth \\
  Department of Music \\
  University of Birmingham \\
  \texttt{c.p.haworth@bham.ac.uk} \\
  \AND
  Xavier Serra \\
  Department of Information and\\
  Communications Technologies \\
  Universitat Pompeu Fabra \\
  \texttt{xavier.serra@upf.edu} \\
  \And
  Rodrigo F. Cádiz \\
  Music Institute and Department\\
  of Electrical Engineering \\
  Pontificia Universidad Católica de Chile \\
  \texttt{rcadiz@uc.cl} \\
}

\makeatletter
\gdef\@aimcthanks{}
\makeatother

\usepackage{xcolor}